\documentclass[a4paper,10pt]{article}
\usepackage{amsmath}
\usepackage{amssymb}

\usepackage{sectsty}
\sectionfont{\fontsize{12}{15}\selectfont}

\usepackage{multicol}
\usepackage[export]{adjustbox}
\usepackage{subcaption}

\usepackage[usenames,dvipsnames]{color}
\usepackage{sansmath}
\usepackage[a4paper,margin=1in,headsep=7.5mm,footnotesep=2.2\baselineskip]{geometry}
\usepackage{times}
\usepackage{newtxtext,newtxmath}
\usepackage{mathtools}
\usepackage{graphicx}
\usepackage{tikz}
\usepackage[Symbol]{upgreek}
\usepackage{accents}
\usepackage{float}
\usepackage{epstopdf}
\usepackage[font=small,labelfont=bf]{caption}
\usepackage{fancyhdr}
\usepackage{changepage}

\DeclareCaptionLabelFormat{adja-page}{\hrulefill\\#1 #2 \emph{(\tit{previous page})}}

\usepackage[hang,flushmargin,bottom]{footmisc} 

\definecolor{armygreen}{rgb}{0.14, 0.71, 0.15}
\definecolor{darkgreen}{rgb}{0.08, 0.48, 0.18}
\definecolor{darkred}{rgb}{0.86, 0.153, 0.153}
\definecolor{azure}{rgb}{0.0, 0.5, 1.0}
\definecolor{bole}{rgb}{0.82, 0.57, 0.22}

\usepackage[numbers]{natbib}
\usepackage[colorlinks=true,
            linkcolor=blue,
            urlcolor=blue,
            citecolor=red]{hyperref}

\definecolor{armygreen}{rgb}{0.14, 0.71, 0.15}
\definecolor{darkgreen}{rgb}{0.08, 0.48, 0.18}
\definecolor{darkred}{rgb}{0.86, 0.153, 0.153}
\definecolor{azure}{rgb}{0.0, 0.5, 1.0}
\definecolor{bole}{rgb}{0.82, 0.57, 0.22}

\def\EatH{Einstein@Home }

\newcommand{\tit}[1]{{\fontfamily{ppl}\selectfont \textit{#1}}}
\newcommand{\qag}[1]{{\fontfamily{qag}\selectfont #1}}

\def\Hz{$\,$Hz}

\def\ccos{\mathsf{cos}}
\def\csin{\mathsf{sin}}
\def\ctan{\mathsf{tan}}

\def\F{\mathcal{F}}

\def\AvTwoF{2{\overline{\mathcal{F}}}}
\def\B{{\hat{\beta}_{\mathsf{S/GLtL}}}}
\newcommand{\Bi}[1]{{\hat{\beta}_{\mathsf{S/GLtL}}^{\,#1}}}

\def\Fspaces{\textsf{F}-\textrm{space }}
\def\Sspaces{\textsf{S}-\textrm{space }}
\def\Fspace{\textsf{F}-\textrm{space}}
\def\Sspace{\textsf{S}-\textrm{space}}

\def\chii{\upchi_i}

\def\chiiF{\upchi_i^{\textsf{F}}}

\def\GammaSeed{\Gamma_\mathsf{S}}
\def\GammaOcc{\Gamma_\mathsf{L}}

\def\RF{\mathcal{R}^\mathsf{F}}
\def\RFik{\mathcal{R}_{i,k}^\mathsf{F}}

\def\RSik{\mathcal{R}_{i,k}^\mathsf{S}}

\def\RFstar{\mathcal{R}_{i}^{\mathsf{F}\ast}}
\def\RSstar{\mathcal{R}_{i}^{\mathsf{S}\ast}}

\def\BFiOne{\mathcal{B}_{i,1}^\mathsf{F}}
\def\BFiOneSquare{[{\BFiOne}]^2}
\def\BFir{\mathcal{B}_{i,r}^\mathsf{F}}
\def\BFirSquare{[\mathcal{B}_{i,r}^\mathsf{F}]^2}

\def\BFirpoSquare{[\mathcal{B}_{i,r+1}^\mathsf{F}]^2}

\def\BSiOne{\mathcal{B}_{i,1}^\mathsf{S}}
\def\BSiOneSquare{[{\BSiOne}]^2}
\def\BSiOneSquare{[\mathcal{B}_{i,1}^\mathsf{S}]^2}
\def\BSir{\mathcal{B}_{i,r}^\mathsf{S}}
\def\BSirSquare{[\mathcal{B}_{i,r}^\mathsf{S}]^2}

\def\BSirpoSquare{[\mathcal{B}_{i,r+1}^\mathsf{S}]^2}
\def\NF{{\mathrm{N}}_\mathsf{F}}
\def\NS{{\mathrm{N}}_\mathsf{S}}
\def\CF{\mathrm{C}_{\mathsf{F}}}
\def\CS{\mathrm{C}_{\mathsf{S}}}
\def\nFiOne{\mathrm{n}_{i,1}^\mathsf{F}}
\def\nFir{\mathrm{n}_{i,r}^\mathsf{F}}

\def\nSiOne{\mathrm{n}_{i,1}^\mathsf{S}}
\def\nSir{\mathrm{n}_{i,r}^\mathsf{S}}
\def\nSirpo{\mathrm{n}_{i,r+1}^\mathsf{S}}
\def\Pthr{\mathsf{P}_\mathsf{th}}
\def\Dthr{\mathsf{D}_\mathsf{th}}
\def\Gthr{\mathsf{G}_\mathsf{th}}
\def\Eff{\mathcal{E}}
\def\dsky{\mathrm{d}_\mathsf{sky}}

\begin{document}
\fancyhead[L]{\footnotesize\tit{Avneet Singh et al}}
\fancyhead[R]{{\footnotesize \tit{published as} \href{https://journals.aps.org/prd/abstract/10.1103/PhysRevD.96.082003}{\tit{\textbf{Physical Review D}} 96(8):082003}}}
\newpage
\topskip15pt
\begin{flushleft}
\textbf{\large Adaptive clustering algorithm for continuous gravitational wave searches}\linebreak

{\small Avneet Singh$^\mathrm{1,\,2,\,3,\color{blue}{\dagger},\color{blue}{\ddagger}}$\let\thefootnote\relax\footnote{$^\mathrm{\color{blue}{\dagger}}$\color{blue}{avneet.singh@aei.mpg.de}; $^\mathrm{\color{blue}{\ddagger}}$ \color{blue}{avneet.singh@ligo.org}}, Maria Alessandra Papa$^\mathrm{1,\,2,\,4}$, Heinz-Bernd Eggenstein$^\mathrm{2,\,3}$, Sin\'ead Walsh$^\mathrm{1,\,2}$
 \linebreak\linebreak}
{\footnotesize $^1$ Max-Planck-Institut f{\"u}r Gravitationphysik, am M{\"u}hlenberg 1, 14476, Potsdam-Golm\\
$^2$ Max-Planck-Institut f{\"u}r Gravitationphysik, Callinstra{$\upbeta$}e 38, 30167, Hannover\\
$^3$ Leibniz Universit{\"a}t Hannover, Welfengarten 1, 30167, Hannover\\
$^4$ University of Wisconsin-Milwaukee, Milwaukee, Wisconsin 53201, USA}
\end{flushleft}
\setcounter{footnote}{0}
\begin{center}
\begin{abstract}
In hierarchical searches for continuous gravitational waves, clustering of candidates is an important post-processing step because it reduces the number of noise candidates that are followed-up at successive stages \citep{fstatGCLSC}\citep{fstatGCMethods}\citep{S6FU}. Previous clustering procedures bundled together nearby candidates ascribing them to the same root cause (be it a signal or a disturbance), based on a predefined cluster volume. In this paper, we present a procedure that adapts the cluster volume to the data itself and checks for consistency of such volume with what is expected from a signal. This significantly improves the noise rejection capabilities at fixed detection threshold, and at fixed computing resources for the follow-up stages, this results in an overall more sensitive search.
This new procedure was employed in the first \EatH search on data from the first science run of the advanced LIGO detectors (O1) \citep{O1AS20-100}. 
\end{abstract}

\end{center}

\begin{multicols}{2}
\section{Introduction}
\label{sec:intro}
In searches for continuous gravitational wave (CW) signals (e.g. \citep{S6BucketStage0, S6FU, S5GC1HF, S6CasA, S5Early, S4Early, S5EHHough, fstatVSR1, fstatGCLSC}), like in many other gravitational wave searches, the detection statistic can be triggered both by signals and by noise disturbances. Furthermore, when the signal or disturbance is strong, it typically does not trigger only a single template waveform but also many nearby ones. 

`Clustering' is the procedure through which we assess elevated detection statistic template points close enough to each other in parameter space that might arise from the same root cause, i.e. signal or noise disturbance. The reason for doing this is that the clustering properties help discriminate candidates due to signals from the candidates due to disturbances, and in certain cases (e.g. loud disturbances), bundle together large numbers of candidates together which one does not need to assess separately. In case of hierarchical sub-threshold searches (e.g. \citep{fstatGCMethods,S6FU}), clustering is performed on the candidates from the first stage. This significantly reduces the number of candidates for subsequent follow-up at fixed threshold on the detection statistic. Hence, at fixed computing budget for the follow-up stages, clustering allows to lower the threshold and increase the sensitivity of the search. 

In previous searches using a clustering procedure, the cluster volume was defined once and for all, based on the average clustering properties of signals \citep{fstatGCMethods, S6FU}. In this paper, we present a clustering method that is adaptive, i.e. it adapts the clustering size in each dimension to the local distribution of candidates in parameter space, and then it requires consistency in clustering among the different dimensions. We have named it AdCl procedure (\tit{Adaptive Clustering Procedure}).

As the name suggests, the AdCl procedure adapts its parameters to the data. If the data were pure Gaussian noise, all this sophistication would not be necessary. Hence, in order to illustrate the AdCl under realistic and relevant conditions, throughout this paper we use small (50 mHz) frequency-domain snippets of data from the first Advanced LIGO observing run (O1).

The paper is organised as follows. In section \ref{sec:general}, we introduce the fundamental idea behind the adaptive clustering procedure; in section \ref{sec:algodetail}, we detail how it functions and introduce the various parameters that characterise it. In section \ref{sec:comp}, we present and compare the performance of this procedure against the clustering procedure used in previous searches. The last section summarises the main findings and discusses prospects.


\section{Clustering of candidates}
\label{sec:general}

A typical all-sky CW search covers the entire sky, a large frequency range and a certain range of spin-down values. In this parameter space, grids are set up and a detection statistic is computed at each grid point.

We indicate a generic grid point with $\uplambda_i\equiv(f_i, \dot{f}_i, \alpha_i, \delta_i)$, with $i=1~... ~\mathrm{N}$, and the detection statistic calculated at that grid point with $\Gamma_i$. Here, $\alpha_i, \delta_i$ are the equatorial sky coordinates of the signal template, while $f_i$ and $\dot{f}_i$ denote the frequency and the first-order spin-down respectively. The result of the search are the ensemble of $\kappa_i\equiv(\uplambda_i,\Gamma_i)$. We concentrate on the subset of these results that are interesting, i.e. where the detection statistic values are elevated above some predefined threshold ($\GammaOcc$). Let's assume that there are $\mathrm{M}$ such results. We will refer to these as the candidates.

Operationally, the clustering procedure is an iterative process and it was first introduced in \citep{fstatGCMethods}: we begin with the highest detection statistic value in our results, corresponding to, say, candidate $\kappa_{i(1)}$, where ``1'' identifies the first iteration of the clustering procedure (i.e. the first cluster). The candidate $\kappa_{i(1)}$ is also called the \tit{seed} for the first cluster. We then find elevated detection statistic values ``nearby'' $\uplambda_i$, and we associate them with $\kappa_{i(1)}$. These set of points will form the first cluster, and they -- along with the seed $\kappa_{i(1)}$ -- will be referred to as the \tit{occupants} of the cluster. We proceed to remove these occupants associated with $\kappa_{i(1)}$ from the original set of candidates. In the next iteration, we consider the highest detection statistic value among the remaining set of candidates, now  $\kappa_{i(2)}$, i.e. the seed for the second cluster. We again find elevated detection statistic values nearby $\kappa_{i(2)}$ and associate them with it. The occupants of the second cluster are again removed from the set of remaining candidates. This process is repeated with $\kappa_{i(3)}$, $\kappa_{i(4)}$, $\kappa_{i(5)}$ and so on. The process ends when we have no more seeds left above a certain predefined detection statistic threshold ($\GammaSeed$).
\end{multicols}
\begin{figure}[H]
\begin{subfigure}{0.5\textwidth}
\includegraphics[width=77.0mm]{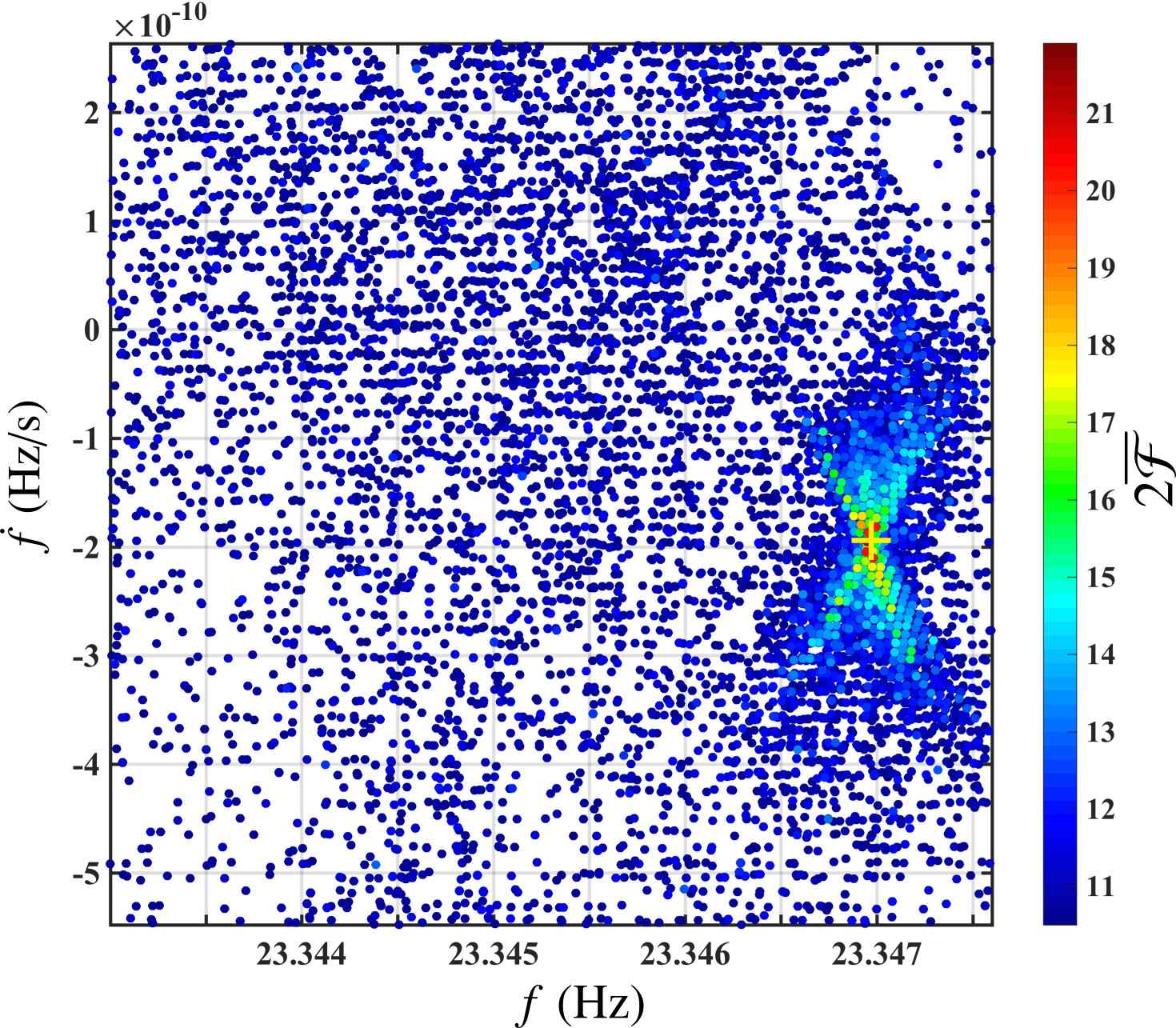}
\end{subfigure}
\begin{subfigure}{0.5\textwidth}
\includegraphics[width=77.0mm]{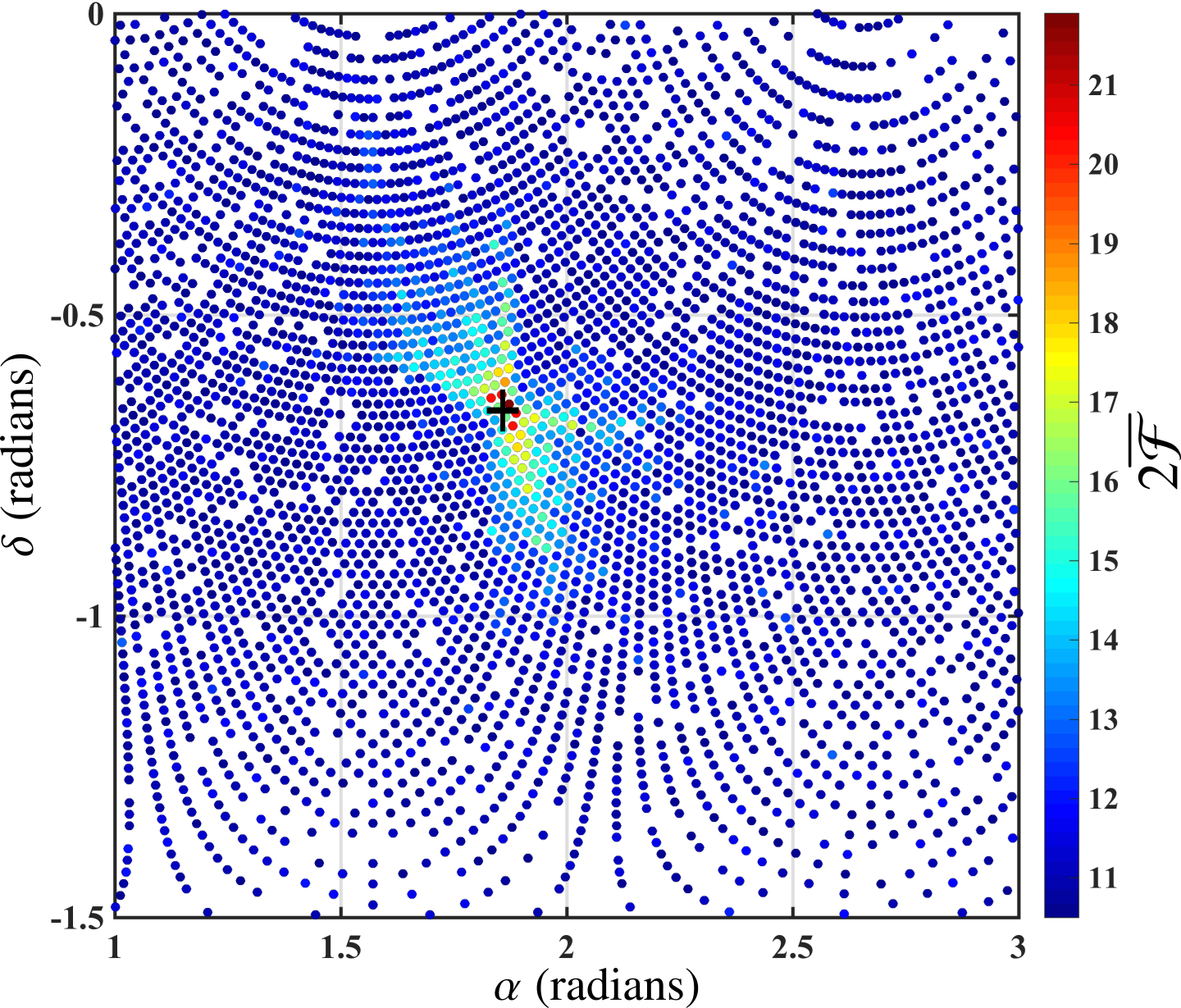}
\end{subfigure}
\caption{{\small \textbf{(\tit{top panel})} Distribution of $\F$-statistic values ($\AvTwoF$) in the parameter space near a fake signal in noise (from LIGO O1 data). Note the elevated $\AvTwoF$ values in the neighborhood of the injection. The elevated $\AvTwoF$ values are clearly coincident in frequency-spindown and the sky. The location of the injection is marked with `+'.}}
\label{fig:example1}
\end{figure}\setcounter{figure}{0}
\begin{figure}[H]
\begin{subfigure}{0.5\textwidth}
\includegraphics[width=77.0mm]{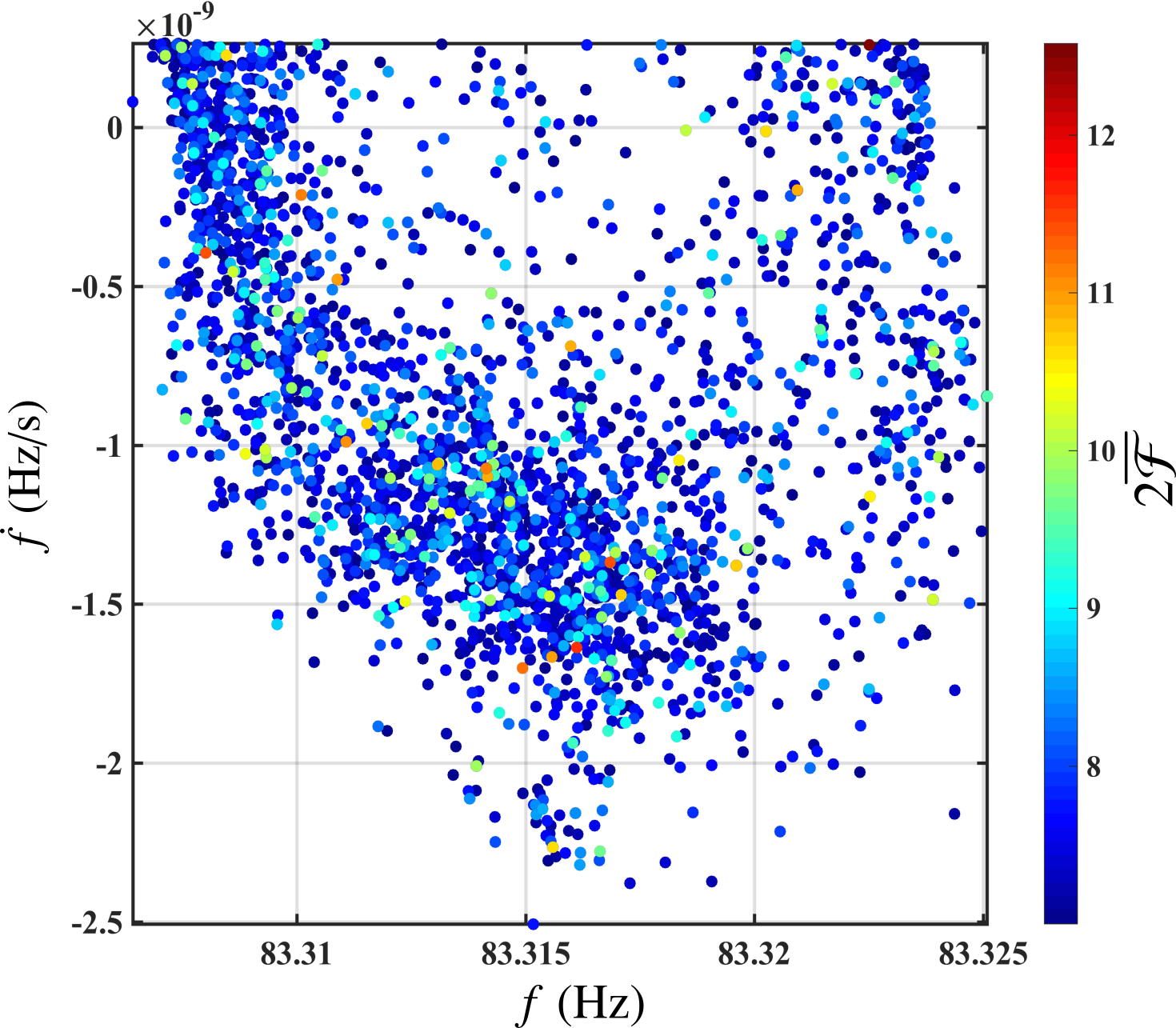}
\end{subfigure}
\begin{subfigure}{0.5\textwidth}
\includegraphics[width=77.0mm]{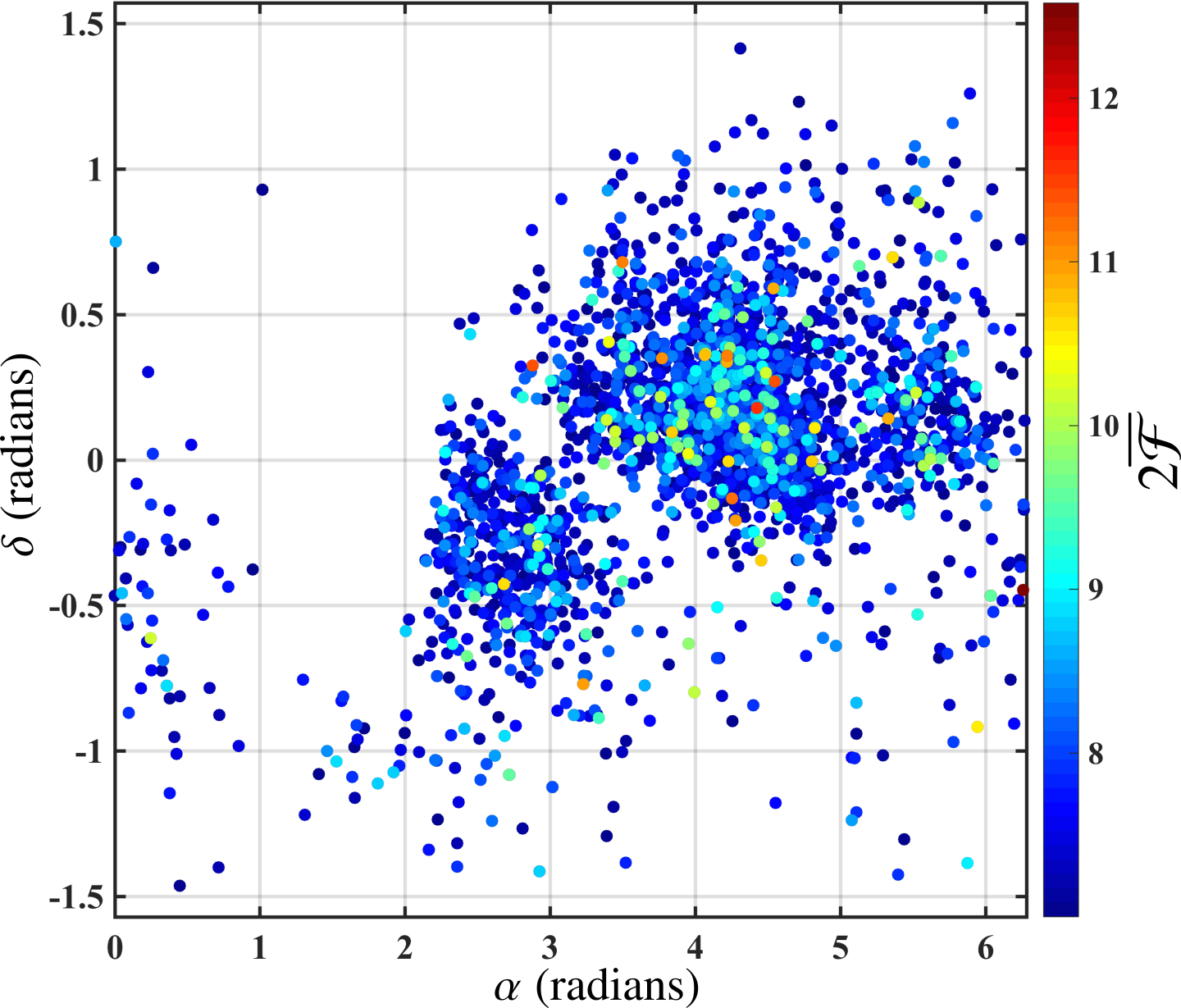}
\end{subfigure}
\caption{{\small \textbf{(\tit{bottom panel})} Distribution of $\mathcal{F}$-statistic values ($\AvTwoF$) in the parameter space in the vicinity of a typical noise disturbance in LIGO O1 data. In contrast with the top panel, the elevated $\AvTwoF$ values due to the disturbance are not coincident between frequency-spindown and sky.}}
\label{fig:example2}
\end{figure}

\begin{multicols}{2}
The core of the AdCl procedure procedure lies in identifying an over-density of candidates in frequency and spin-down around each seed, in determining its extension, and in checking whether that set of candidates also presents an over-density in sky around its seed. These features are trademark signatures of signals (e.g. see figure \ref{fig:example1} top panel), and they are not shared by most noise fluctuations/disturbances (e.g. see figure \ref{fig:example2} bottom panel). We note that previous clustering algorithms did not require such coincident over-densities between frequency-spindown and the sky, and hence, lead to a higher number of false alarms.

Furthermore, the AdCl procedure dynamically defines the clustering neighbourhood based on the data itself. In contrast, previous procedures derived a static clustering neighbourhood around the seed based on average clustering properties of the signals independently of the data. Thus, the AdCl procedure enables us to bundle together any over-density that extends over large volumes of parameter space as a single follow-up candidate, and hence, the number of candidates to follow-up from highly populated parameter space regions decreases significantly.   

\section{The cluster size}
\label{sec:algodetail}
\subsection{A measure of distance in frequency and spin-down space (\boldmath{\Fspace})}
\label{sec:metric}

The clustering is applied to a set of candidates $\upchi_1$ whose detection statistic value is above a certain threshold $\GammaOcc$; Let's assume that there are M such candidates:
\begin{equation}
\upchi_1 := \{\kappa_\ell\}\indent\mid\;\;\Gamma_{\ell}\geq\GammaOcc,
\label{eq:band}
\end{equation}
where $1 \leq \ell \leq \mathrm{M}$. 

In general, at each iteration $i$, the clustering procedure defines a new cluster, and it does this by operating on a set of candidates $\upchi_{i}$. We indicate the seed for the $i$-th cluster with $\kappa_{\ell(i)}$, with $\ell(i)$ being the index that corresponds to the candidate with the loudest detection statistic value among the candidates in $\upchi_{i}$. We constrain the cluster seed to exceed a fixed threshold $\GammaSeed$, which in general is larger than $\GammaOcc$. The clustering procedure stops at iteration $\mathrm{N}_\mathsf{c}+1$ when there are no more candidates with detection statistic values above  $\GammaSeed$ in  $\upchi_{\mathrm{N}_\mathsf{c}+1}$, i.e. when $\Gamma_{\ell(\mathrm{N}_\mathsf{c}+1)}< \GammaSeed$. 

At each iteration $i$, we define as Euclidean distance $\RF_{i,k}$ in frequency and spin-down space (\Fspace) between the cluster seed $\kappa_{\ell(i)}$ and every other candidate $\kappa_k$ in $\upchi_{i}$: 
\begin{equation}
\RF_{i,k}:=\sqrt{\Bigg[\frac{f_k-f_{\ell(i)}}{\delta\! f}\Bigg]^2 + \Bigg[\frac{\dot{f}_k-\dot{f}_{\ell(i)}}{\delta\!\dot{f}}\Bigg]^2} \;\;\forall\;\;\kappa_k\in\upchi_{i},
\label{eq:Fdistance}
\end{equation}
where, $\delta\!f$ and $\delta\! \dot{f}$ are the frequency and spin-down grid spacings used in the search. Note that at fixed $\RF_{i,k}$, \eqref{eq:Fdistance} is an ellipse in \textsf{F}-space centered at $(f_{\ell(i)},\dot{f}_{\ell(i)})$ and with axes of half-length $\delta\!{f}\times\RF_{i,k}$ and $\delta\!\dot{f}\times\RF_{i,k}$.

\subsection{Distribution of distances in \textsf{F}-space}
\label{sec:distrFdistances}
We define the cluster radius for the $i$-th cluster based on the distribution of the distances $\RFik$ in \textsf{F}-space. In order to derive such a distribution, we must bin the distances $\RFik$ appropriately. 

\begin{figure}[H]
\includegraphics[width=80.00mm]{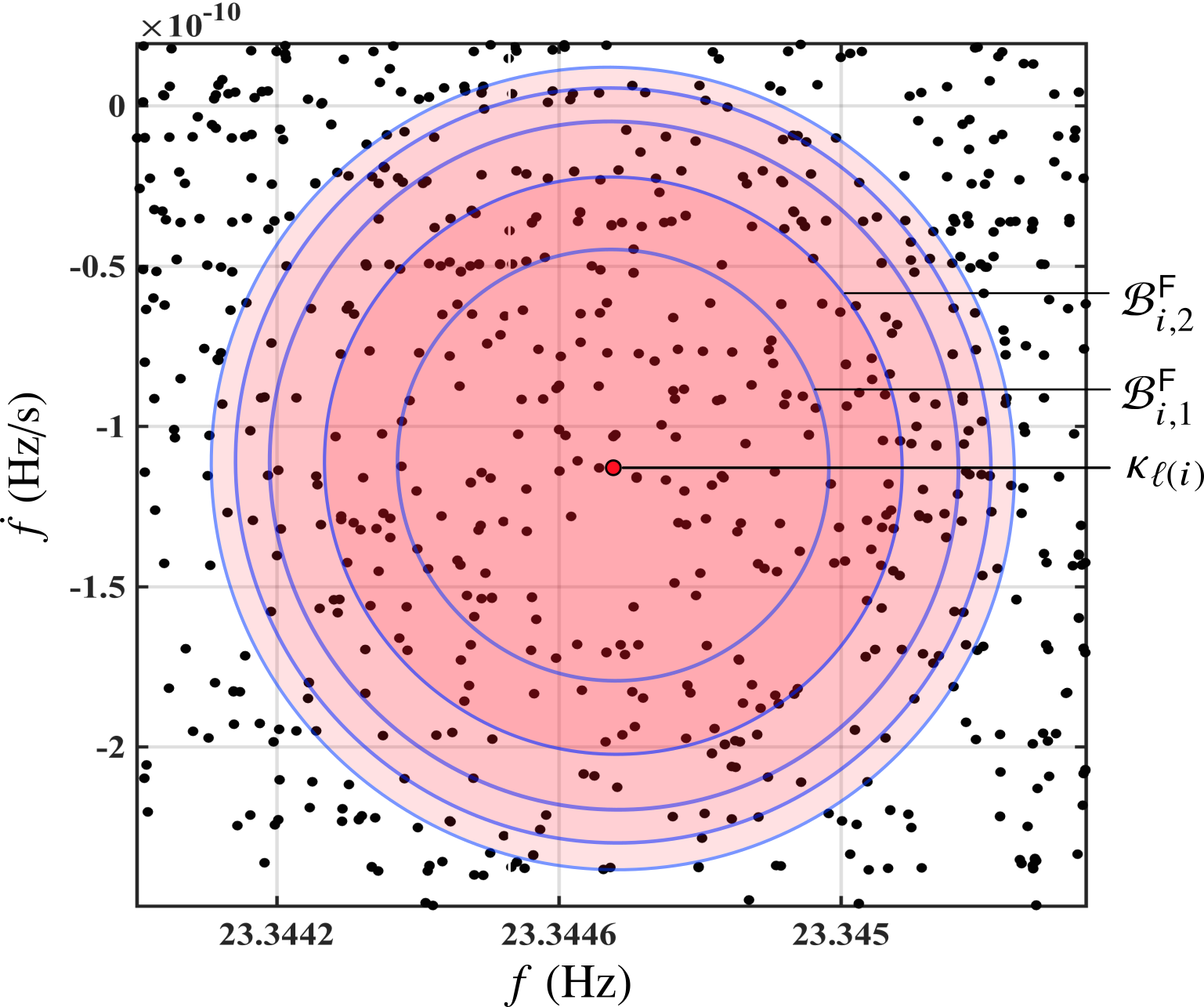}
\caption{{\small Example of annular binning in \textsf{F}-space, defined by \eqref{eq:BiOne}--\eqref{eq:FGridEx}. The values of the parameters are: $\delta\! f=8.3\times 10^{-7}$\Hz, $\delta\!\dot{f} = 1.3\times10^{-13}$\Hz/s. An ad-hoc value for $\BFiOne$ of $1.2\times10^3$ ($\NF = 50$) is taken. The area within each annuli is constant, hence the annuli get thinner with distance from the center. Note that the figure shows only the first 5 annuli for clarity; in total, there are $\NF^2 = 2500$ such annuli.}}
\label{fig:annuli}
\end{figure}

The binning in \textsf{F}-space naturally takes the form of concentric elliptical annuli ($f^r,\dot{f^r}$) at distances $\BFir$ from the seed. The index $r$ denotes the different bins. The edges ($f^1,\dot{f^1}$) of the first bin are defined by the equation
\begin{equation}
\sqrt{\Bigg[\frac{f^1-f_{\ell(i)}}{\BFiOne~\delta\! f}\Bigg]^2 + \Bigg[\frac{\dot{f}^1-\dot{f}_{\ell(i)}}{\BFiOne~\delta\!\dot{f}}\Bigg]^2} = 1.
\label{eq:BiOne}
\end{equation}
The successive bins are defined by the recursive relation
\begin{equation}
\BFirpoSquare - \BFirSquare =\BFiOneSquare ~~~\textrm{for all bins } r = 1,2,3~...~,
\label{eq:FGrid}
\end{equation} 
which requires that the area of the annuli is constant and equal to $ \uppi \BFiOneSquare$ (see figure \ref{fig:annuli}). Note that each annulus encloses an equal number of parameter space points. The relation \eqref{eq:FGrid} can be explicitly solved to yield
\begin{equation}
\BFir = \sqrt{r}\,\BFiOne ~~~\textrm{for all bins } r = 1,2,3~...~,
\label{eq:FGridEx}
\end{equation} 

The value of $\BFiOne$ is chosen as 
\begin{equation}
\label{eq:RFbin}
\BFiOne=\displaystyle\frac{1}{\NF}\mathcal{R}^\mathsf{F}_{i,\,\mathsf{max}} ~~~{\textrm{with}}~~~\mathcal{R}^\mathsf{F}_{i,\,\mathsf{max}}=\max_k\{\RFik\}.
\end{equation}
Comparing \eqref{eq:RFbin} with \eqref{eq:FGridEx} and setting $\mathcal{R}^\mathsf{F}_{i,\,\mathsf{max}} = \BFir$, we find that $\NF^2$ is equal to the total number of $r$ bins. $\NF$ is the determined as follows: the candidate count $\nFir$ in the various $r$ bins is determined for a test value of $\NF$, say ${\NF}_t$.  If the condition,
\begin{equation}
\nFiOne({\NF}_t)\geq\CF~\langle\nFir({\NF}_t)\rangle\;\;\text{where}\;\; r = 1,2,3~...~,
\label{eq:dFrac}
\end{equation}
is not satisfied, we iteratively decrease ${\NF}_t$ by one (${\NF}_t \rightarrow {\NF}_t - 1$) until \eqref{eq:dFrac} is verified, and set $\NF={\NF}_t$. In \eqref{eq:dFrac}, the angled brackets indicate the average over the $r$ bins, and $\CF>1$. Note that ${\NF}_t$ should be large enough such that $\BFiOne$ is comparable with the signal containment region in \Fspaces \citep{S6FU}. Further, $\CF$ encodes the over-density requirement, and for low amplitude signals, this requirement is very lax: $\CF{\gtrsim}1$, which means that the procedure picks the finest binning for which we at least do not have an under-density around the seed.

We note that in a sub-threshold search, the clustering procedures are in principle sensitive to the parameter $\GammaOcc$: the over-densities of signal candidates due to a weak signal will only be observable down to certain detection statistic values, below which the density of noise candidates will be high enough that the over-density due to the signal candidates will not be appreciable. The threshold $\GammaOcc$ could, in principle, be optimally placed at the level just above when this effect begins to take place. However, this is difficult to determine. By setting $\CF\gtrsim 1$, we appreciate the smallest over-density possible, and hence, ease the dependency of the procedure on $\GammaOcc$.

If for some $i$-th cluster, no resolution (no $\NF$ value) can be found that meets the requirement of \eqref{eq:dFrac}, then only the seed $\kappa_{\ell(i)}$ is removed from $\upchi_{i}$ and the resulting set of candidates defines $\upchi_{i+1}$. The $i$-th cluster, $\phi_i$, is classified as a \tit{single-occupant-cluster}. 

\begin{figure}[H]
\centering\includegraphics[width=78.0mm]{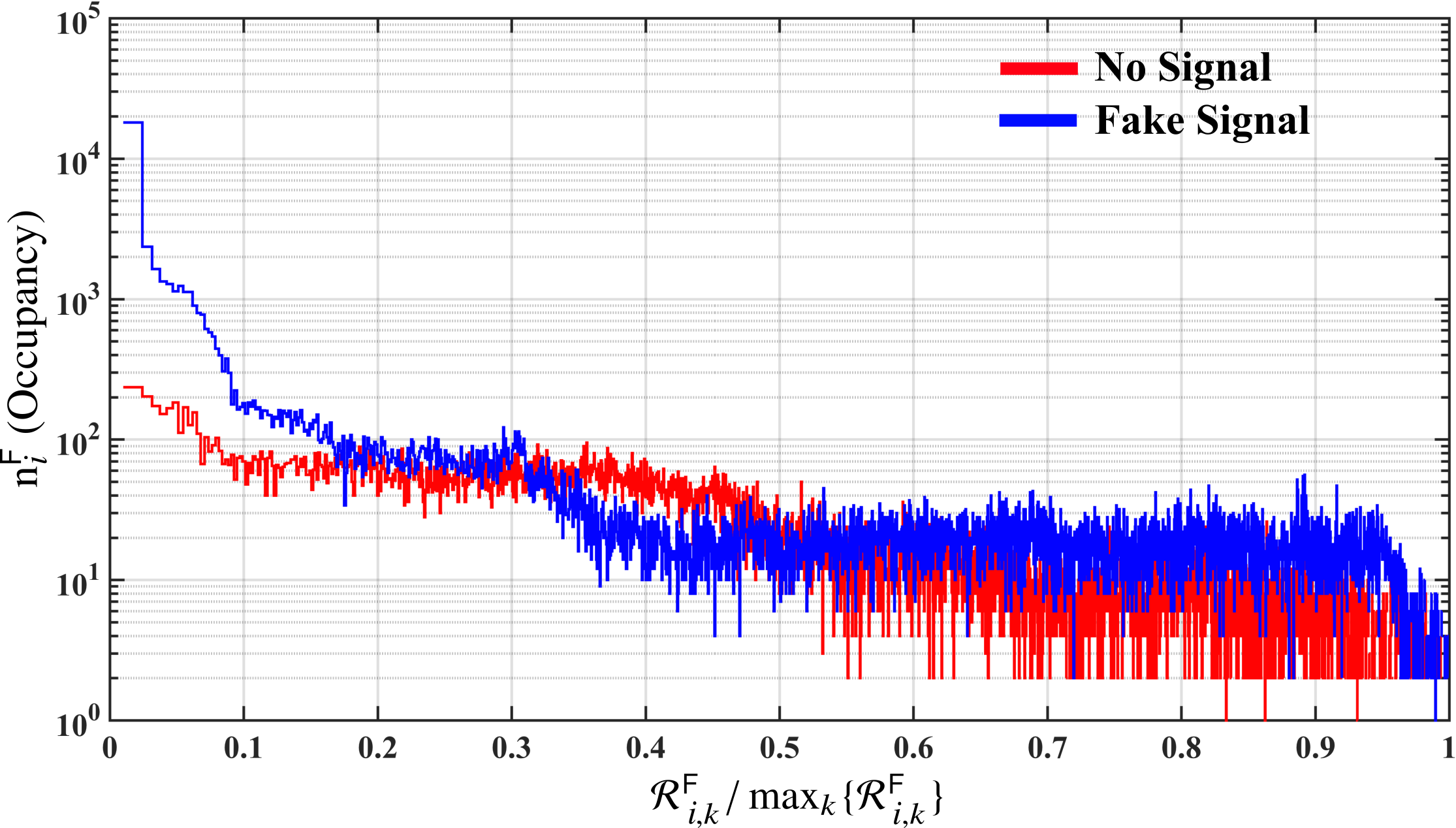}
\caption{{\small Distribution of $\RFik$ for a noise-only data-set (red) and for a data-set also containing a fake signal (blue). The grid spacing $\BFiOne$ in \textsf{F}-space is defined by $\NF = 50$.}}
\label{fig:comparison}
\end{figure}

In figure \ref{fig:comparison}, we compare the distribution of $\RFik$ values from searches ran on noise data (blue curve), and on fake noise plus a CW signal (red curve). The red distribution presents a clear maximum near the seed $\kappa_{\ell(i)}$, i.e. there is an evident over-density of candidates near the seed. We want to estimate the extent of this over-density, and cluster the candidates that form this over-density together.

\subsection{Cluster size in \Fspace}
\label{sec:clusterF}

For every $i$-th cluster, we use the distribution of distances $\RFik$ in order to determine the cluster radius in \Fspace: the cluster radius $\RFstar$ is the value of the distance defined in \eqref{eq:Fdistance} at which we have the first local minimum of $\nFir$. Since the $\RFik$ histogram has typically many fluctuations, in order to estimate more accurately the position of the first minimum of the underlying distribution, we determine its shape with a fitting procedure which smoothens out the random fluctuations. 

We fit the data $\nFir$ in two stages. In the first stage, the data is separately fitted with two functions $\mathsf{G}$ (a superposition of Gaussians) and $\mathsf{S}$ (a superposition of sinusoids):
\begin{equation}
\begin{multlined}
\mathsf{G}(x) = \sum_{l=1}^{\mathrm{m}_1}\mathsf{G}_{l}(x);~~~\mathsf{S}(x) = \sum_{l=1}^{\mathrm{m}_2}\mathsf{S}_{l}(x).
\label{eq:Fit1}
\end{multlined}
\end{equation}
This step is implemented using a compiled \qag{\small {MATLAB}} executable (using the package \tit{fit}), which provides support for $\mathrm{m}_1,\mathrm{m}_2\in[1,8]$. For each fit, we choose the highest value of $\mathrm{m}_1$ and $\mathrm{m}_2$ that is able to fit the data within the standard tolerances defined by the program. The fitted curves $\mathsf{G}$ and $\mathsf{S}$ are summed and re-normalised, and the output is then fit again with a Gaussian function, yielding $g_{i}^\mathsf{F}$. This second fit smoothens out the small scale fluctuations and leaves us with a clear view of the over-densities in \Fspace.
\begin{figure}[H]
\centering\includegraphics[width=78.0mm]{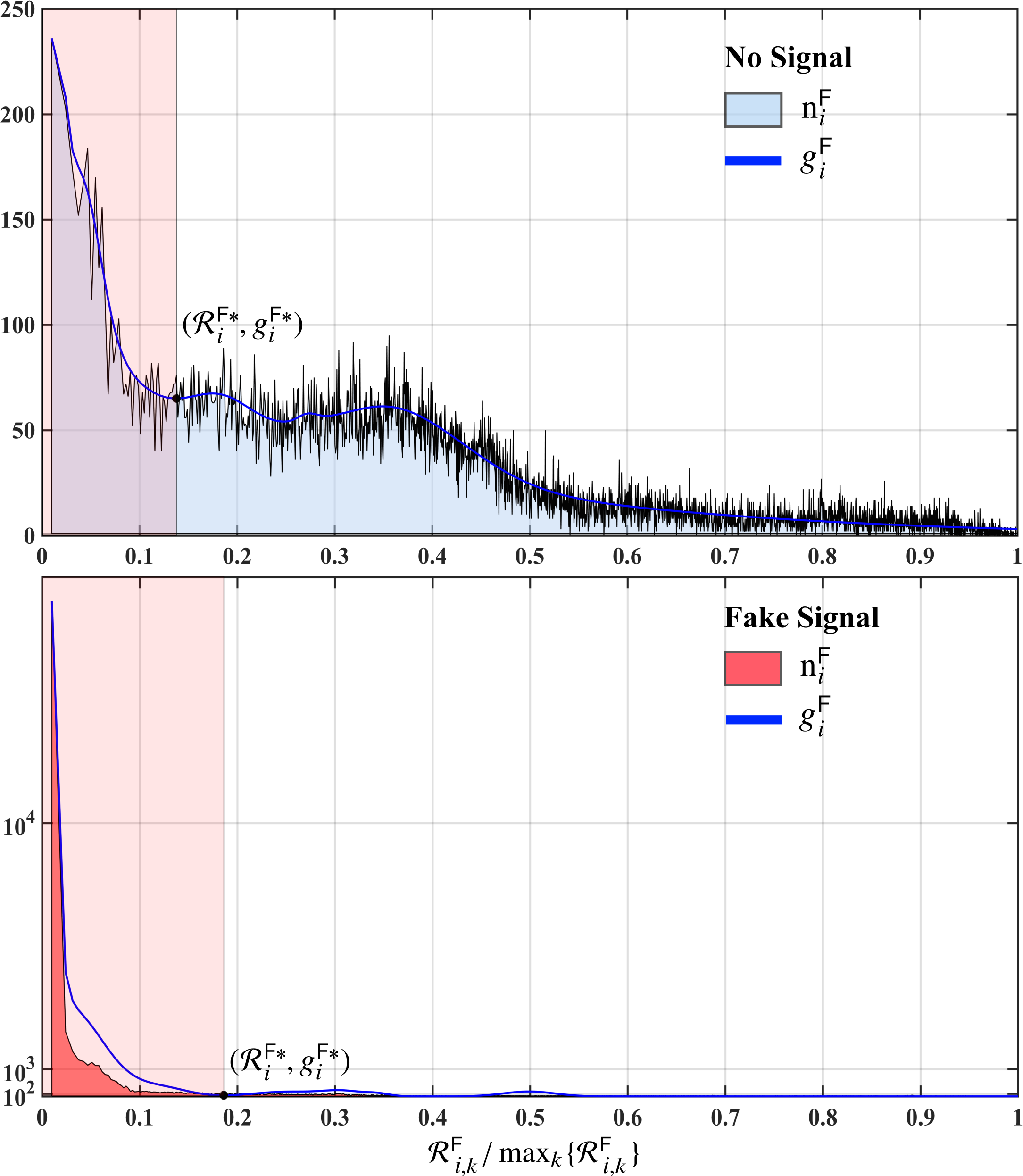}
\caption{{\small Fit to the distribution of $\RFik$ for a noise-only data-set (top panel) and for a data-set also containing a fake signal (bottom panel). The grid spacing $\BFiOne$ is defined by $\NF = 50$. We can see that the fitting procedure contours the shape of the distribution while ignoring small scale fluctuations, and returns a good measure of the over-density.}}
\label{fig:fit}
\end{figure}

Finally, we can identify the local minimum of $g_{i}^\mathsf{F}$ closest to the origin and take that as the radius $\RFstar$ in \Fspaces of the $i$-th cluster. In figure \ref{fig:fit}, we show an example of the fitting procedure on purely noise data (top panel) and in data containing a fake signal (bottom panel). 

\subsection{Hill parameters in \Fspaces and further constraints}
\label{sec:hill}
The distribution of candidates in parameter space is very diverse, depending on the nature of the noise in the data. Because of this, even an adaptive procedure, such as the one described above, may still generate clusters that spuriously assemble together candidates that are actually independent. In order to counter this, instead of setting more stringent criteria, for example a higher threshold $\CF$, it is more effective to produce a first estimate of the cluster based on liberal parameters, and then scrutinize its topological properties in detail, and further accept, discard or modify the cluster based on these.. The topological properties that we consider for a cluster $i$ are the so-called ``hill parameters''\footnote{We adopt the notion of hill parameters from the concept of `topographic prominence' used in topography/geography, e.g. see \citep{hill}.} \tit{prominence} $\mathsf{P}_{i}$, \tit{dominance} $\mathsf{D}_{i}$ and \tit{goodness} $\mathsf{G}_{i}$:
\begin{equation}
\begin{multlined}
\mathsf{P}_{i} :=\frac{\RFstar}{\displaystyle\max_{k}\{\RFik\}},\label{eq:prom}
\end{multlined}
\end{equation}
\begin{equation}
\begin{multlined}
\mathsf{D}_{i} :=\frac{g_{i}^\mathsf{F}(\BFiOne) - g_{i}^\mathsf{F}(\RFstar)}{g_{i}^\mathsf{F}(\BFiOne)},\label{eq:dom}
\end{multlined}
\end{equation}
\begin{equation}
\begin{multlined}
\mathsf{G}_{i} :=\frac{|\nFiOne - g_{i}^\mathsf{F}(\BFiOne)|}{\nFiOne + g_{i}^\mathsf{F}(\BFiOne)}.\label{eq:good}
\end{multlined}
\end{equation}
The cluster candidates from the set $\upchi_{i}$ are further inspected to check if:
\begin{equation}
\begin{multlined}
\mathsf{P}_{i} \leq \Pthr, ~~\mathsf{D}_{i} \geq \Dthr, ~~ \mathsf{G}_{i} \leq \Gthr.\label{eq:hillpeak}
\end{multlined}
\end{equation}
These thresholds ($\Pthr$, $\Dthr$, $\Gthr$) on the hill parameters restrict the topology of clusters: $\Pthr$ restricts the fraction of the available parameter space that the cluster occupies; $\Dthr$ bounds the contrast between the density of candidates near the seed and at the cluster edge; $\Gthr$ specifies the minimum agreement between the fitted curve and the observed density near the seed. The nature and the values of the constraints in \eqref{eq:hillpeak} is such that they exclude clusters that extend too far in the \textsf{F}-space, and at the same time, show very little contrast with respect to the local noise background; thus we shortlist the kind of clusters that we typically expect from signals.

When a cluster in \Fspaces fails to meet any of the criteria given by \eqref{eq:hillpeak}, we shortlist candidates $\nFiOne$ from the distribution that fall within the first bin $\BFiOne$ around the seed and discard all other candidates from the iteration. This is equivalent to resetting $\RFstar=\BFiOne$. This choice is justified because the failing of the hill parameters means that the shortlisted cluster is not topologically consistent with what we require from a cluster of that extent. However, the initial over-density still remains near the seed and it might be due to a low amplitude signal. In this regard, we do not discard the whole cluster. On the other hand, if the criteria in \eqref{eq:hillpeak} are met, we shortlist all the candidates, including the seed, that fall within our estimated cluster radius $\RFstar$, and discard all other candidates outside the cluster radius.

The candidates clustered in \Fspaces will constitute the $\upchi_{i}^{\mathsf{sky}}$ set and their clustering properties in the sky will be considered further.

\subsection{A measure of distance in the sky (\Sspace)}
\label{sec:clusterS}

We now want to determine whether the shortlisted candidates in $\upchi_{i}^{\mathsf{sky}}$ show any over-density in sky around the seed. If any over-density is found, the candidates constituting this over-density will form the final $i$-th cluster.

As in \Fspace, for each candidate $\kappa_k\in\upchi^{\mathsf{sky}}_{i}$, we introduce a distance in the sky, $\RSik$, to the seed of the $i$-th cluster under consideration: 
\begin{equation}
\RSik:=\sqrt{[{x_k-x_{\ell(i)}}]^2 + [{y_k-y_{\ell(i)}}]^2} \;\;\forall\;\;\kappa_k\in\upchi^{\mathsf{sky}}_{i}.\label{eq:metricSky}
\end{equation}
This definition is justified when the search grids are uniform on some plane ($x, y$), for example the ecliptic plane (e.g. see \citep{S6FU}) or the equatorial plane (e.g. see \citep{S6BucketStage0}). The transformation equations between the sky coordinates ($\alpha,\delta$) and ($x,y$) for a uniform grid on ecliptic plane, are:
\begin{equation}
\begin{cases}
~x={\ccos}\uplambda\,{\ccos}\upbeta\\
~y={\csin}\uplambda\,{\ccos}\upbeta,
\label{eq:xtrans}
\end{cases}
\end{equation}
with 
\begin{equation}
\begin{cases}
~\uplambda = {\ctan}^{-1}\Bigg[\displaystyle\frac{{\csin}\alpha\,{\ccos}\varphi + {\ctan}\delta\,{\csin}\varphi}{{\ccos}\alpha}\Bigg] \\
~\upbeta = {\csin}^{-1}[{\csin}\delta\,{\ccos}\varphi - {\csin}\alpha\,{\ccos}\delta\,{\csin}\varphi].
\label{eq:betatrans}
\end{cases}
\end{equation}
In the expressions above, $\varphi = 23.4^\mathsf{o}$ is the angle of obliquity of the ecliptic with respect to the celestial equatorial plane\footnote{Note that in \eqref{eq:betatrans}, $\uplambda$ must be translated to its correct quadrant by adding $180^\mathsf{o}$ or subtracting $180^\mathsf{o}$.}. The ecliptic plane represents the \Sspaces after this transformation.

\subsection{Distribution of distances in \Sspace}
\label{sec:gridSky}

The binning of the $\RSik$ values is performed in a similar fashion as previously done in \Fspace. The edges of the bins, labeled by $r$, of the $i$-th cluster, satisfy the following relation:
\begin{equation}
\BSirpoSquare - \BSirSquare =\BSiOneSquare ~~~\textrm{for all bins } r = 1,2,3~...~.
\label{eq:RSGrid}
\end{equation}
This recursive relation describes concentric circular annuli in the $(x,y)$ plane enclosing equal areas; the annuli naturally get thinner as we move away from the seed, as shown in figure \ref{fig:annuli}. The first bin is a circle and its area is proportional to $\BSiOneSquare$.

$\BSiOne$ is chosen based on the clustering properties of signals. Precisely, it will depend on the 99\% containment region of the search \citep{S6FU}. This region defines a neighbourhood around a cluster seed originating from a signal, within which the true signal parameters are contained with 99\% confidence. If we indicate with $\dsky$ the width of the search pixels in the $(x,y)$ plane \citep{S6BucketStage0,S6FU}, and with $\mathrm{N}^{99\%}$ the diameter of the 99\% containment region expressed in number of pixels, then we can express $\mathcal{B}_{i,1}$ as
\begin{equation}
\BSiOne:={{\mathrm{N}^{99\%} + \NS}\over 2} ~\dsky,
\label{eq:Bi1}
\end{equation}
where, $\NS$ is a parameter that has to be tuned as shown in section \ref{sec:comp}. Further, \eqref{eq:Bi1} says that the first bin in the sky, i.e. the circle with radius $\BSiOne$, contains all sky pixels within the 99\% containment region, plus (or minus) a tuning term $\NS$. We continue binning recursively according to \eqref{eq:RSGrid} until the width of the bin becomes smaller than a sky-grid pixel; all candidates lying more distant than that point are ignored.

\begin{figure}[H]
\centering\includegraphics[width=78.0mm]{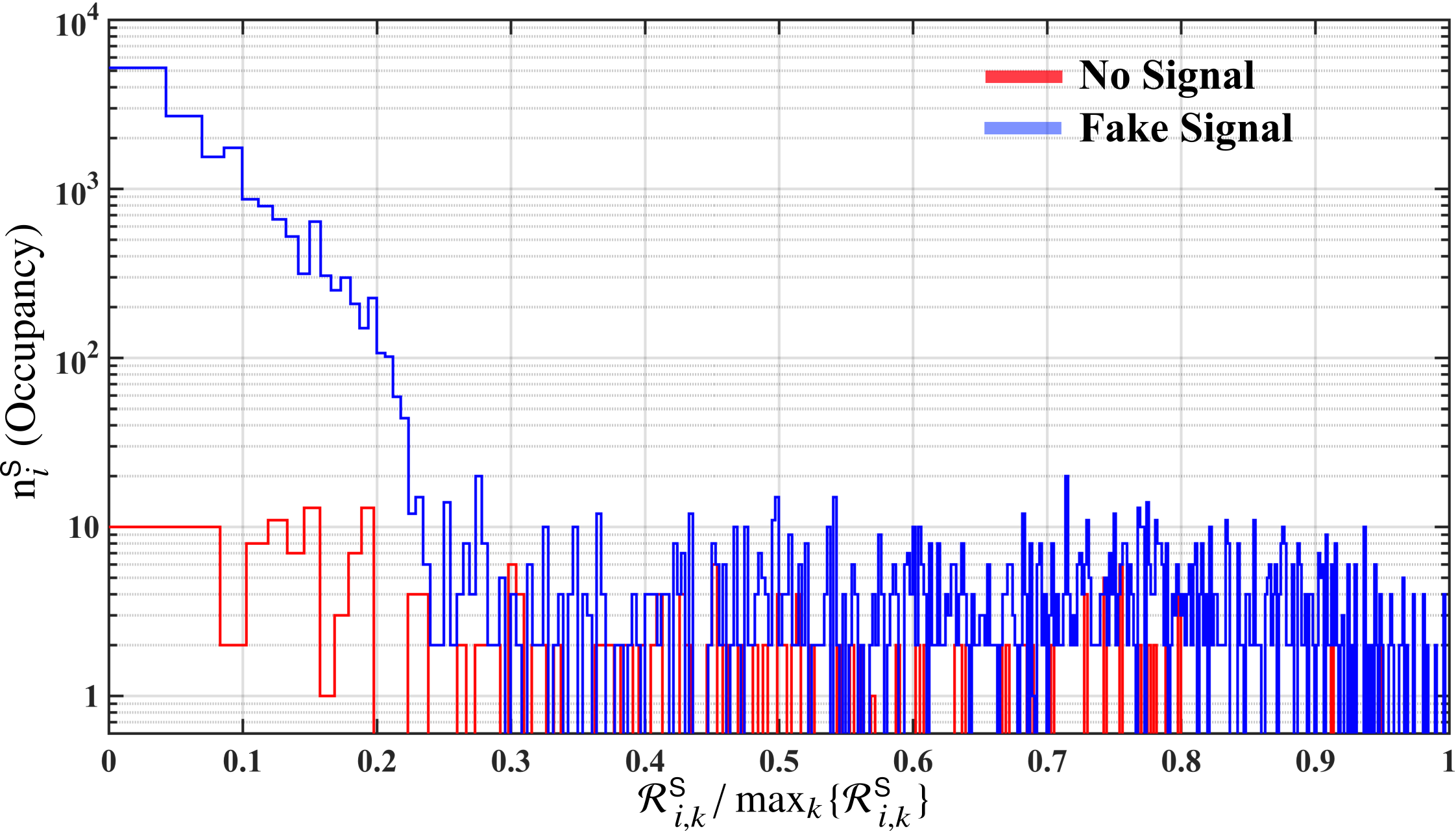}
\caption{{\small Distribution of $\RSik$ values for signal versus noise: $\BSiOne$ for the noise-only case is defined by $\NS = 0$, while for signal it is set to $\NS = 6$.}}
\label{fig:DistSky}
\end{figure}

We find that in disturbed data that contains a large number of noise outliers, a single value of $\NS$ independent of the loudness $\Gamma_i$ of the cluster seed under consideration, makes this clustering procedure very slow. The reason is that very large values of $\Gamma_i$ are often associated with many candidates clustered in \Fspace (highly populated $\chiiF$) that are distributed almost isotropically in the sky. In this situation, if the resolution in the sky ($\BSiOne$) is high, the sky-clustering step eliminates one candidate at the time as a single-occupant-cluster, and this is very inefficient. The solution is to decrease the resolution (increase $\BSiOne$ by increasing $\NS$) with $\Gamma_i$.

In figure \ref{fig:DistSky}, we show the re-normalised distribution of $\RSik$ for a fake signal and LIGO O1 noise. 

\subsection{Cluster in \Sspace}
\label{sec:clusterSky}
In order to estimate the cluster radius in \Sspace, we check for over-densities by analysing the distribution of $\RSik$. 

If the first bin is the most highly populated (i.e. $\nSiOne = \max_r\{\nSir\}$), all the candidates contained within a distance $\RSstar$ are clustered together: 
\begin{equation}
\RSstar=\min_r\Bigg\{\BSir:\frac{{\nSir}-{\nSirpo}}{{\nSir}}>\CS\Bigg\}.\label{eq:testSky}
\end{equation}
$\RSstar$ is the smallest distance at which we have a relative drop in the density of candidates above a certain threshold $\CS$. All candidates within $\RSstar$ constitute, together with the seed, the final $i$-th cluster, $\phi_i$. The set of candidates considered for the next clustering iteration is $\upchi_{i+1}=\upchi_i - \phi_i$.

\begin{figure}[H]
\centering\includegraphics[width=78.0mm]{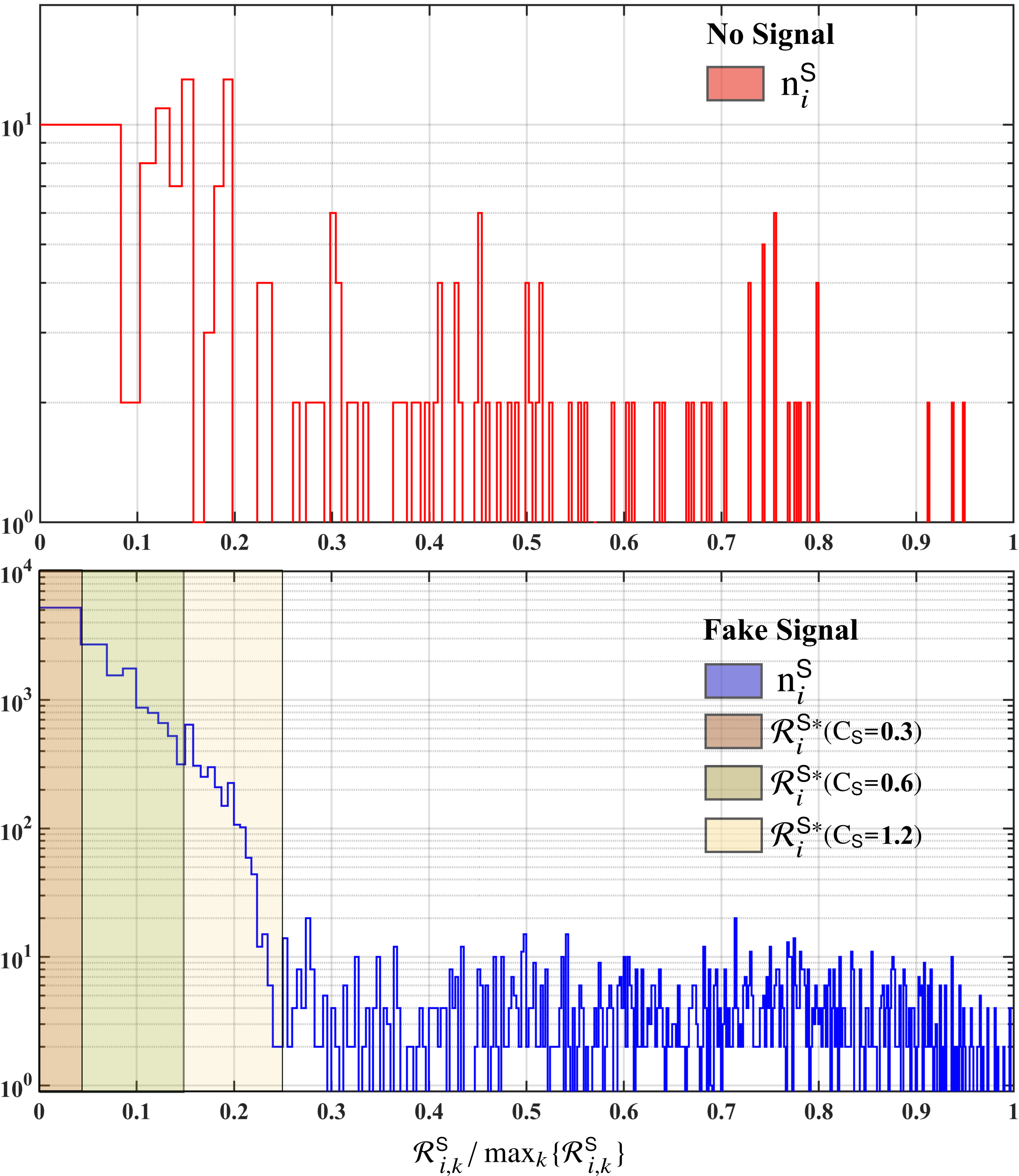}
\caption{{\small Clustering properties in the sky for a data-set containing a signal (lower panel) and a noise data-set (upper panel). The values of $\NS$ are 6 and 0, respectively for the lower and upper panels. Note that the first bin in the pure noise case (top panel) is not the most highly populated, and hence failed the clustering criteria ($\nSiOne \neq \max_r\{\nSir\}$), irrespective of the value of $\mathrm{C}_\mathsf{S}$.}}
\label{fig:skycut}
\end{figure}

The value of $\CS$ is chosen based on the localisation properties of signals and leaning on the conservative side, i.e. toward lower values of $\CS$. For instance, in the bottom panel of figure \ref{fig:skycut}, we see that $\RSstar(1.2) > \RSstar(0.6) > \RSstar(0.3)$. Indeed, the lower value of $\CS$ clusters less candidates, but the candidates excluded at this iteration will likely form their own separate cluster at the next iteration. If this second set of points were due a signal, with a lower $\CS$, they would be associated to the correct seed.

If the first bin is not the most highly populated, the final cluster $\phi_i$ will contain only the seed $\kappa_{\ell(i)}$. All the other candidates remain un-clustered, and available for association with another cluster in the set $\upchi_{i+1} = \upchi_i - \kappa_{\ell(i)}$. 

This recursive procedure continues until there are no more candidate seeds, i.e. no more candidates with detection statistic value above the threshold $\GammaSeed$. In figure \ref{fig:samplerun1}, \ref{fig:samplerun2} in appendix A, we show a snapshot of the procedure for the first iteration on data $\upchi_1$ for a fake signal and near-Gaussian noise. 

\section{Performance}
\label{sec:comp}

\begin{table}[H]
\begin{center}
\bgroup
\def\arraystretch{1.2}
\begin{tabular}{|l|l|}
\hline
\hline
\textbf{Quantity} & \textbf{Value}\\
\hline
\hline
$\mathrm{T}_{\mathsf{obs}}$ & 4 months \\ \hline
$\mathrm{T}_{\mathsf{coh}}$ & 210 hours\\ \hline
$\mathrm{N}_{\mathsf{seg}}$ & 12       \\ \hline
$\delta\!f$ & $8.3\times 10^{-7}$ Hz   \\ \hline
$\delta\!\dot{f}$ & $1.3\times 10^{-13}$ Hz/s         \\  \hline
$\mathrm{d}_\mathsf{sky}$ ($f=100${\Hz}) &  20 arcmin \\  
\hline
\hline
\end{tabular}
\egroup
\end{center}
\caption{The clustering procedure is applied to the output from this all-sky search.}
\label{table:params}
\end{table}

We characterise the performance of the AdCl procedure and compare it with the old clustering procedure, used in \citet{S6FU}. We show how the tuning parameters were chosen in an actual search \citep{O1AS20-100}, with parameters given in table \ref{table:params}.

The two clustering procedures are compared at the same value of seed threshold $\GammaSeed$, and with the other parameters optimally tuned. 

\subsection{Clustering parameters}
\label{sec:tuningParams}
We will consider two different data inputs to the clustering procedure, one suitable for a high-significance search (loud signals), and the other for a sub-threshold search (weak signals). In the former search, the detection statistic is $\AvTwoF$, the corresponding $\GammaSeed$ and $\GammaOcc$ thresholds are 12.0 and 10.5 respectively, and the value of $\NS$ for the $i$-th cluster is:
\begin{equation}
\NS(\AvTwoF_i)=
\begin{cases}
0 & {\textrm{if}}~~~\AvTwoF_i < 18\\
\AvTwoF_i -18 & {\textrm{if}}~~~18 \leq \AvTwoF_i \leq48 \\
30 & {\textrm{if}}~~~\AvTwoF_i > 48. 
\label{eq:NeFstat}
\end{cases}
\end{equation}
Alternatively, for the second search, the detection statistic is the line- and the transient line-robust statistic $\B$ \citep{DavidTransients, O1AS20-100}, the corresponding $\GammaSeed$ and $\GammaOcc$ thresholds are 5.5 and 4, respectively, and $\NS$ for the $i$-th cluster is:
\begin{equation}
\NS(\Bi i)=
\begin{cases}
0 & {\textrm{if}}~~~\Bi i < 15\\
\Bi i -15 & {\textrm{if}}~~~15 \leq \Bi i \leq 35 \\
31 & {\textrm{if}}~~~\Bi i > 35. 
\label{eq:NeBsgl}
\end{cases}
\end{equation}
This is the set-up appropriate for a search like \citep{O1AS20-100}. 

The reason why we consider searches with different detection statistics is historical: at the time when we started characterising the AdCl procedure, we were planning to use it for a high-significance search on quiet bands, as done in \citep{S6BucketStage0}. In this case, the simplest detection statistic to use is $\AvTwoF$, and all the false alarm and detection efficiency studies were performed with this statistic. It was only later that we realised that the quality of the data in the low-frequency range was such that a high-significance search was not possible: we would have many candidates above threshold, and we would have to carry out a large scale follow-up. Due to these complications, for this search, the use of the $\B$ was necessary. In the absence of large disturbances, the empirical relationship between the two detection statistics is $\AvTwoF \equiv 0.419~\B + 10.855 $.

The other parameters are chosen as described in the previous sections and they are equal for both types of searches, and their values are:
\begin{equation}
\begin{cases}
\NF\in [25,50], \CF=1.2\\
\CS=0.25 \\
\Pthr=0.25, \Dthr=0.05, \Gthr=0.1.
\label{eq:otherParams}
\end{cases}
\end{equation}

On the other hand, the old clustering uses a fixed cluster size corresponding to the 99\% containment regions in the various dimensions. In case of the high-threshold $\AvTwoF$ search:
\begin{equation}
\begin{multlined}
\AvTwoF\equiv
\begin{cases}
\Delta f=1.15\times 10^{-4} \text{\Hz},\\ \Delta\dot f = 5.6\times 10^{-11} \text{\Hz/s},\\ \Delta^{\textsf{sky}}=6\times 6 \textrm{ sky-pixels},
\end{cases}
\end{multlined}
\label{eq:oldparam1}
\end{equation}
while for the sub-threshold $\B$ search:
\begin{equation}
\begin{multlined}
\B \equiv
\begin{cases}
\Delta f=1.85\times 10^{-4} \text{\Hz},\\ \Delta\dot f = 8.5\times 10^{-11} \text{\Hz/s},\\ \Delta^{\textsf{sky}}=9\times 9 \textrm{ sky-pixels}.
\end{cases}
\end{multlined}
\label{eq:oldparam2}
\end{equation}

\subsection{Safety}
\label{sec:safety}

Naturally, the clustering procedure needs to be safe, i.e. it should not discard real signals. Thus, we choose the clustering parameters to yield the lowest false alarm rate for a very low false dismissal rate. We now show the detection efficiencies for the clustering parameters outlined in \eqref{eq:NeFstat}, \eqref{eq:NeBsgl}, \eqref{eq:otherParams}.

We estimate the detection efficiency by performing Monte-Carlo simulations of gravitational wave signals in real data taken from the LIGO O1 run. By using the real LIGO data instead of fake Gaussian noise, we derive a realistic benchmark of the performance. In a nutshell, the fake signals are added to the real data, the search is run, and the clustering procedure is applied.

\begin{figure}[H]
\centering\includegraphics[width=78.0mm]{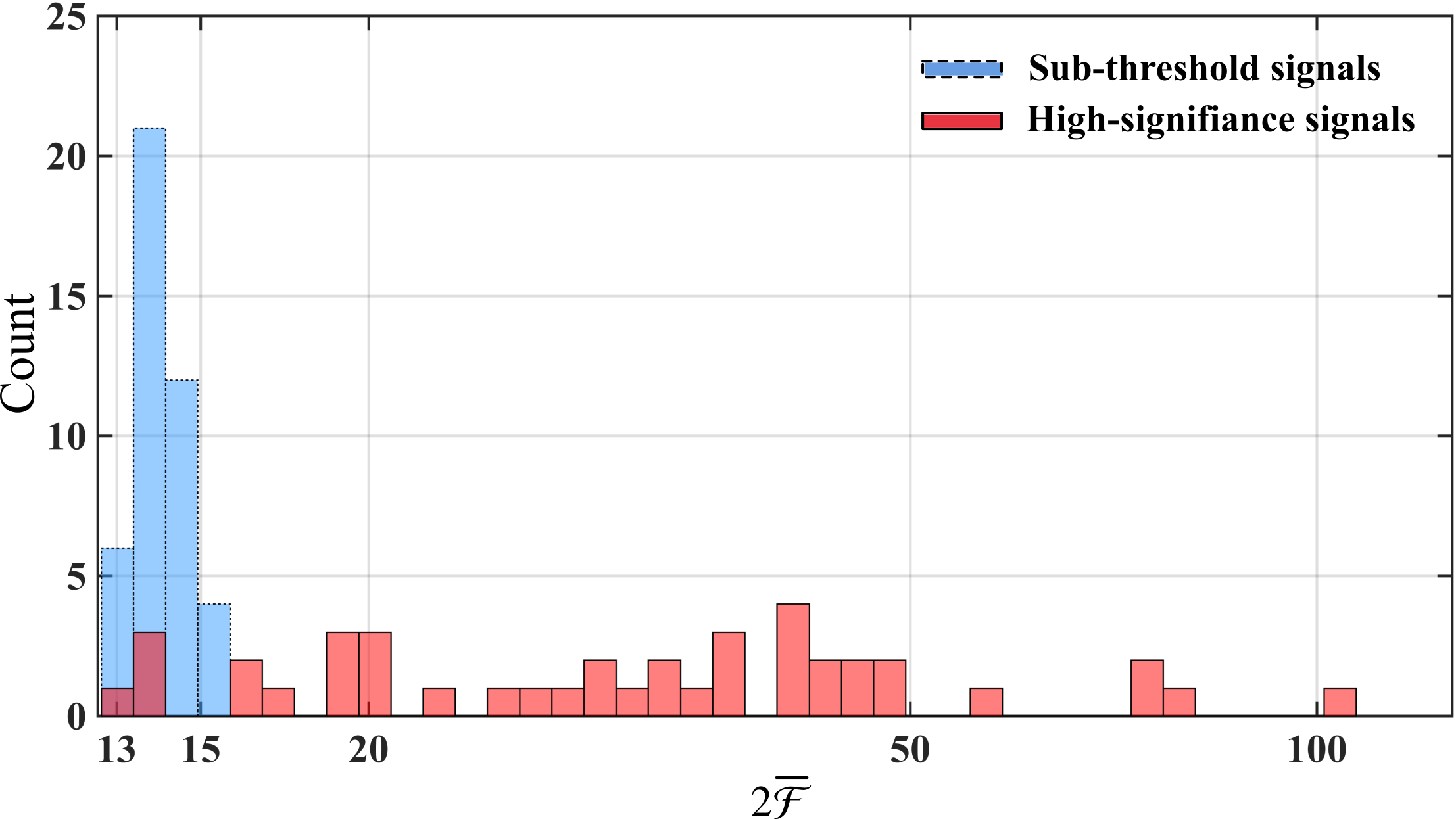}
\centering\includegraphics[width=78.0mm]{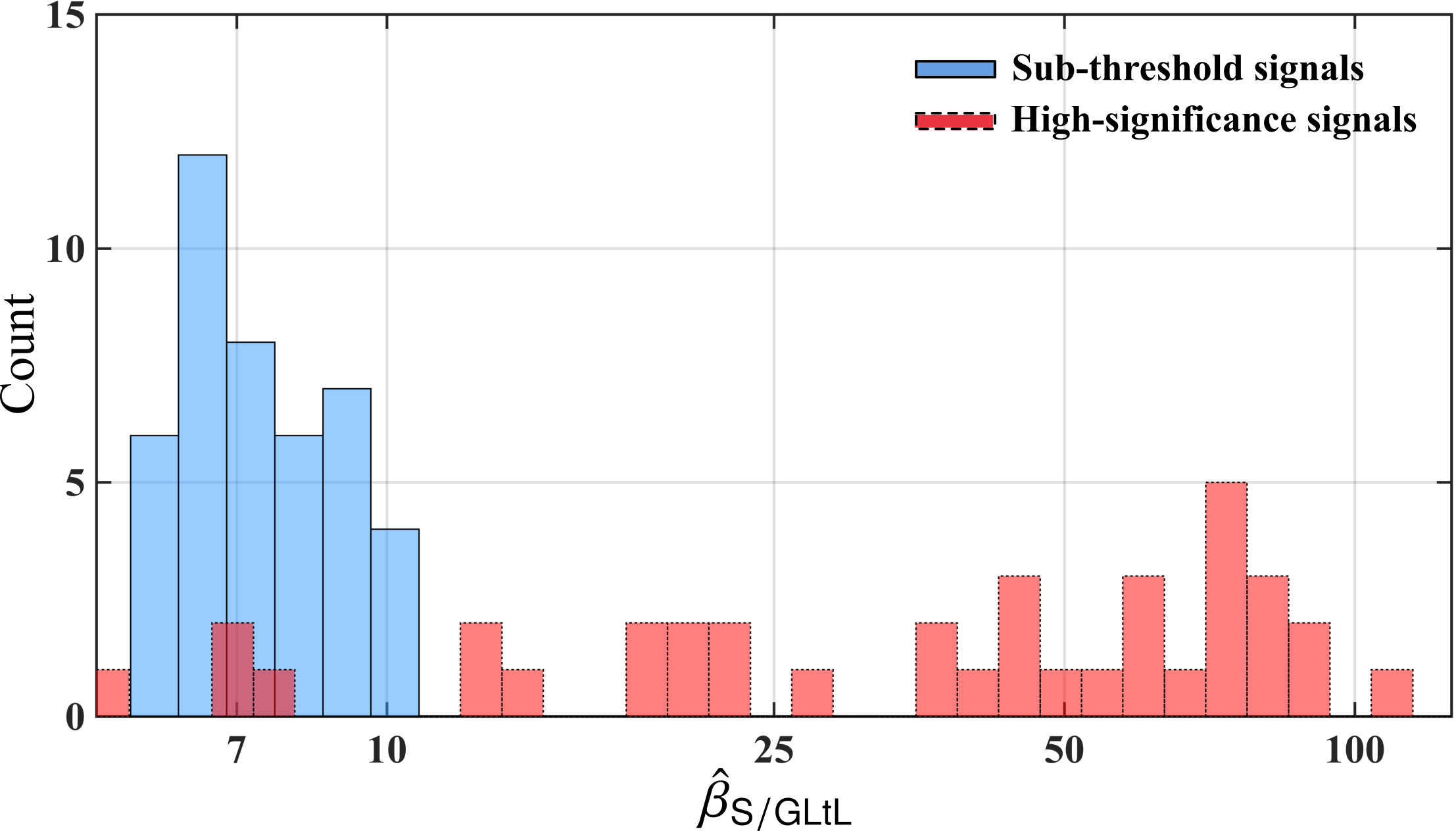}
\caption{{\small Distribution of the values of the detection statistics of the sub-threshold and high-significance signals added to the data to characterize the performance of the clustering procedure. Note that the high-significance signal population is not purely high-significance; it also contains a few signals at low values of the detection statistic (less than $10\%$ below $\AvTwoF = 14.0\equiv\B = 7.5$). Meanwhile, the sub-threshold search may be considered as purely sub-threshold (none above $\B = 10.5\equiv\AvTwoF = 15.3$).}}
\label{fig:InjTwoF}
\end{figure}

The population of signals have parameters reasonably uniformly distributed in frequency, spin-down and sky-position, and with amplitudes that yield the detection statistic values shown in figure \ref{fig:InjTwoF}.

The detection efficiency $\Eff$ is defined as the ratio of the number of candidates from signals recovered by the clustering procedure with the total number of signals with detection statistic value above $\GammaSeed$. For a signal to be recovered by the clustering procedure, we require that the signal parameters lie within the 99\% containment region of the seed parameters (we remind the reader that the detection statistic value of the seed must also exceed $\GammaSeed$). This means that if there were a follow-up stage on the cluster seeds, the true signal parameters would lie within the searched region, and if there were no follow-ups, the signal parameters would lie within the quoted parameter uncertainties.

\subsection{Noise Rejection}
\label{sec:FAR}

We estimate the false alarm rate by applying the clustering procedure to the same search output data as described in the previous section, just without fake signals. 

The input to the clustering procedure are $\mathcal{N}_{\textsf{in}}$ candidates, with detection statistic values greater than $\GammaSeed$. At the output of the clustering procedure, we have $\mathcal{N}_{\textsf{out}}$ candidates. We define the noise reduction factor $\textrm{NR}$ as:
\begin{equation}
\textrm{NR} := 1 - {\mathcal{N}_{\textsf{out}} \over \mathcal{N}_{\textsf{in}}} ~~~~ {\textrm{on noise}}.
\label{eq:noisereduction}
\end{equation}
Naturally, $0\leq{\textrm{NR}}\leq 1$, and higher values of NR denote lower number of noise candidates after the clustering procedure. 
\end{multicols}
\begin{table}[H]
\begin{center}
\bgroup
\def\arraystretch{1.2}
\begin{tabular}{|c|c|c|c|}
\hline
\hline
&  & \textbf{AdCl Procedure} & \textbf{Old Procedure}\\
\hline
\hline
\textbf{High-significance} & NR & 65.9{\%} & $\leq$ 40.0{\%} \\
\cline{2-4}
\boldmath{$\AvTwoF$} \textbf{search} & $\Eff$ & 97.6{\%} & 95.1{\%} \\
\hline\hline
\textbf{Sub-threshold} & NR & 90.5{\%} & $\leq$ 74.1{\%} \\
\cline{2-4}
\boldmath{$\B$} \textbf{search} & $\Eff$ & 95.5{\%} & $>95.0{\%}$ \\
\hline
\hline
\end{tabular}
\egroup
\end{center}
\caption{Comparison of the noise rejection (NR) and the detection efficiences ($\Eff$) of high-significance and sub-threshold searches between the new and old clusturing procedures.}
\label{table:PerformanceTable}
\end{table}
\begin{multicols}{2}
\subsection{Results}
\label{sec:results}
The performance results for the AdCl and the old clustering procedures are shown in table \ref{table:PerformanceTable}. For a high-significance search, the detection efficiency, exceeding 95\%, is high for both the procedures, but the new clustering has a noise rejection (NR) which is significantly higher (nearly 66\% versus 40\%) than the one achieved by the previous method. 

In a sub-threshold search, we set a low enough threshold on the detection statistic of the seed ($\GammaSeed$) such that we expect a large number of candidates to exceed this limit, just due to random noise. The underlying idea behind this is that with successive follow-up stages, one is able to weed out the noise and identify a signal that, at the first stage of the hierarchy, was hidden by a multitude of false alarms. In this regime, the clustering procedure operates in an environment of the most uniformly and densely populated candidates. The signal signature used by the clustering procedure are local over-densities around the cluster seed, coincident in \Fspaces and \Sspace. But the cluster seed is, at every iteration $i$, the loudest candidate in the set $\chii$, and when the signal is weak, i.e. its amplitude is comparable to the amplitude of many of the candidates, it might not be picked as a seed. For this reason, the detection efficiency is lower for a sub-threshold search with respect to a high-threshold search. In order to compare the performance of AdCl procedure with the old procedure, we fix the detection efficiency at >$\,95{\%}$ by lowering $\GammaOcc$ to 3.4 for the old procedure (keeping $\GammaOcc = 4.0$ for AdCl procedure). In this case, the AdCl procedure improves the noise rejection (NR) by 22{\%} over the old procedure.

The results of table \ref{table:PerformanceTable} refer to signal-frequency bands where the data is fairly uniformly distributed in parameter space, i.e. there are no extended regions of the parameter space that host enhanced values of the detection statistic, as in the case of the top panel of figure \ref{fig:example1}. 
Moreover, the AdCl procedure performs very well in disturbed conditions, and this is important because the disturbed regions typically yield a lot of spurious candidates. 

In noisy regions, the new clustering procedure has a NR of 98.9\%, compared to $\leq$ 91.1\% for the old procedure in a $\AvTwoF$ search. We expect similar results for noisy data in a $\B$ search. The NR values in the disturbed bands are higher than those in quiet bands because each cluster comprises more candidates above $\GammaSeed$ in noisy bands than in the quiet bands. This is expected merely due to higher density of disturbances. The new clustering procedure has a higher NR than the old method because it adapts the cluster size to the local over-density and can get as big (or small) as it needs, in order to accommodate the features in the data.

A rigorous quantitative assessment of the detection efficiency in disturbed bands is hard to make because the results would depend not only on the location of the fake signals in parameter space but also their numbers with respect to the disturbances. In such scenarios, there is no unbiased way to pick the fake signal population. However, based on the fact that for a cluster to be identified, we only require a seed above threshold and concurrent clustering around that seed in both \Fspaces and in \Sspace, we do not expect the presence of more candidates due to disturbances (which generally do not cluster in the parameter space) to interfere too much with the identification of the signal clusters. On the contrary, the old procedure does not require a local over-density around the seed and it might happen that a signal candidate gets associated with a higher random fluctuation; this cluster may not satisfy the over-density criteria in the AdCl procedure which may have led to a wrong estimation of the follow-up region. Thus, by requiring the seed to be centered at a local over-density, the new procedure avoids this type of occurrence. This might slightly favour the detection efficiency of the AdCl procedure with respect to the old one.

\section{Conclusions}
\label{sec:conc}

The clustering procedure that we propose in this paper is more effective at reducing the number of candidates to be considered in follow-up stages while achieving comparable, if not better, detection efficiency with respect to the procedure used in previous searches.  Since we operate at fixed computing budget, the number of candidates that a given follow-up stage can search, is fixed. Hence, a higher noise rejection means a lower detection threshold. In a search like the Einstein@Home O1 low-frequency search \citep{O1AS20-100}, the new clustering has allowed us to lower the $\B$ threshold. In disturbed bands, the noise rejection is even higher.  

There are two main reasons for the observed improvements. The first reason is that the AdCl procedure is more demanding than the old one, i.e. a cluster has to display a more pronounced over-density of candidates compared to nearby noise. The second reason is that, since the cluster size is estimated on the data itself, the clustering algorithm adapts itself to it and is capable of bundling together a large number of candidates arising from extended regions of parameter space. 

Another advantage of the AdCl procedure compared to the old one is that, by relying on local over-densities of candidates, the false alarm rate does not increase with a decreasing value of the threshold $\GammaOcc$, which is significantly necessary for low-significance searches. 

However, the AdCl procedure may well go through many iterations before discarding a single candidate as a single occupancy cluster and restoring the rest of the candidates for future consideration. This, especially in noisy bands, can make it rather slow. The variable sky-binning depending on the seed amplitude is a way to ease this issue, and quite certainly, further use will inspire other ways to make the procedure faster in all noise conditions.

Currently, the tuning parameters ($\NF$, $\CF$, $\NS$, $\CS$) and the hill parameters ($\Pthr$, $\Dthr$, $\Gthr$) are chosen to represent the approximate topology of the clusters that we expect from signals. These approximate values are chosen upon visual inspection of the fake signals injected in LIGO data at many values of the signal amplitude $h_0$. To improve the estimates on these parameters, one would require to perform a much larger number of Monte-Carlo simulations (in $\uplambda$ and $h_0$), and then estimate the cluster properties. This remains a difficult task due to limited computational resources and very large parameter space of the tuning and hill parameters. The modeling of clusters arising from non-Gaussian noise (such as instrumental artifacts) is even more difficult, especially for unknown sources of disturbances. In principle, this modeling could however help better discern between signals and noise.

There are certainly other possible ways to perform adaptive clustering. One of the methods is to employ machine learning. Besides that, one could also perform more complex parameter space correlation studies of the detection statistic values, similar to the studies done for cosmic microwave background (\textsf{CMB}) surveys \citep{CMB} and large scale structure (\textsf{LSS}) surveys \citep{LSS}. However, such complex analysis methods require much cleaner data, and they are certainly an overkill for the current data-sets.

\section{Acknowledgements}
\label{sec:acknowledgements}
This procedure was used in \citep{O1AS20-100}, and we thank Sergey Klimenko and Evan Goetz for the review of the application of this new clustering procedure to the results of that search. M A Papa and S Walsh gratefully acknowledge the support from \textsf{NSF PHY Grant} {\small \qag{1104902}}. All computational work for this search was carried out on the ATLAS super-computing cluster at the Max-Planck-Institut f{\"u}r Gravitationsphysik, Hannover and Leibniz Universit{\"a}t Hannover. This document has LIGO DCC number {\small \qag{P1700123}}. The implementation of the AdCl algorithm is in progress under the application name {\textsf{lalapps}\_\textsf{AdaptiveClustering}\_\textsf{v1}} in the {{\textsf{lalapps}}/{\textsf{src}}/{\textsf{pulsar}}/} repository.

\begin{center}
$$\ast\ast\ast$$
\end{center}

\bibliographystyle{plainnat}

\end{multicols}
\vspace{10pt}
\begin{multicols}{2}
\section*{APPENDIX}
\label{Appendix}
\subsection*{A: The first cluster}
\begin{figure}[H]
\centering
\includegraphics[width=1.00\columnwidth]{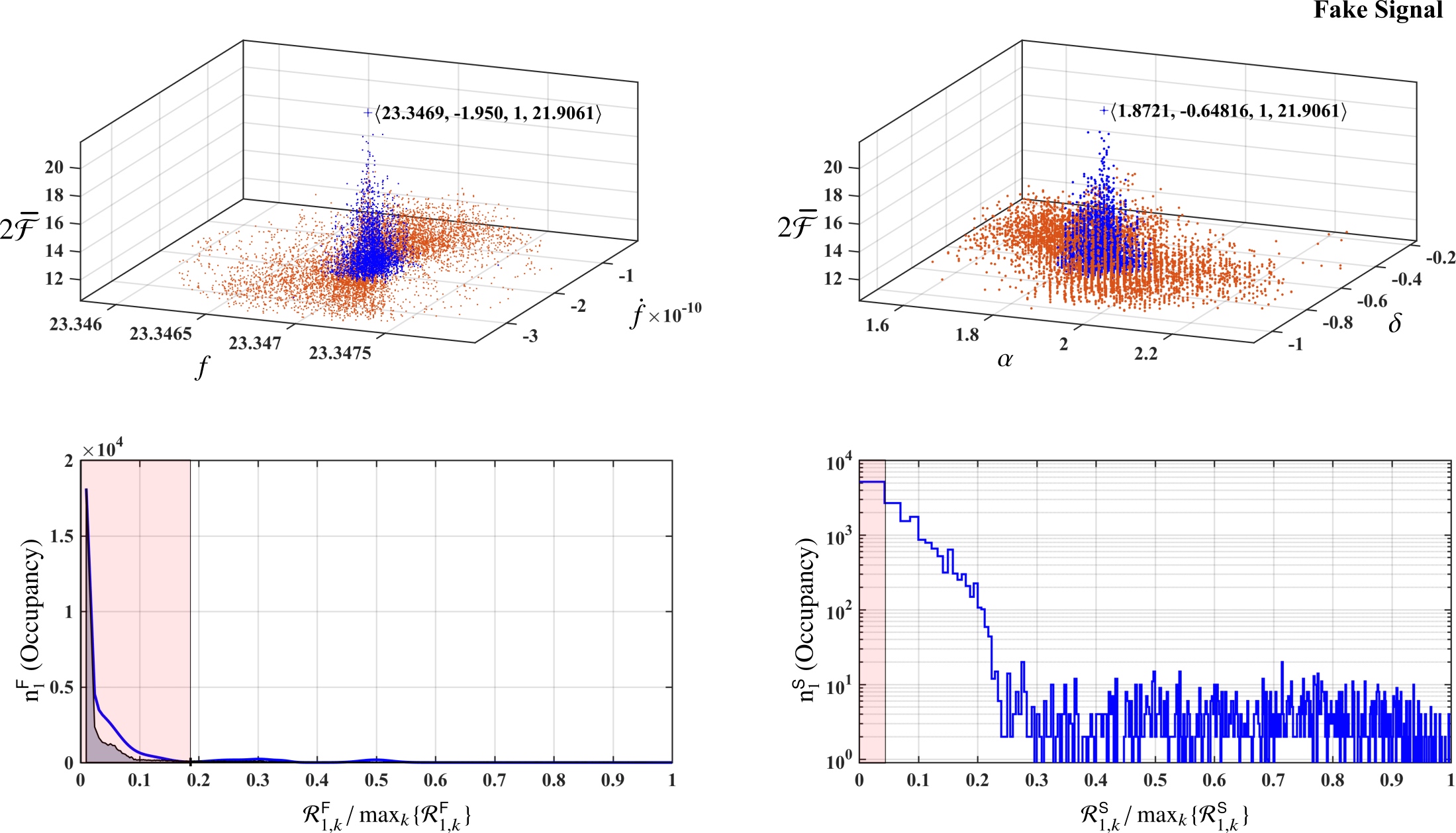}
\caption{(\tit{\textbf{Signal case}}) The orange points are the candidates in $\upchi_1^{\mathsf{sky}}$; the blue points are the subset of these that form the final cluster $\upphi_1$. The corresponding distributions for $\RFik$ and $\RSik$ are shown in the second row of plots. The shaded regions extend up to $\RFstar$ (left plot) and $\RSstar$ (right plot). The seed is marked with a `+'. The numbers in the brackets by the `+' denote: $\langle f, \dot{f}, i, \Gamma_i\rangle$ in \Fspaces and $\langle \alpha, \delta, i, \Gamma_i\rangle$ in \Sspaces, where $\Gamma\equiv \AvTwoF$.}
\label{fig:samplerun1}
\end{figure}

\begin{figure}[H]
\centering
\includegraphics[width=1.00\columnwidth]{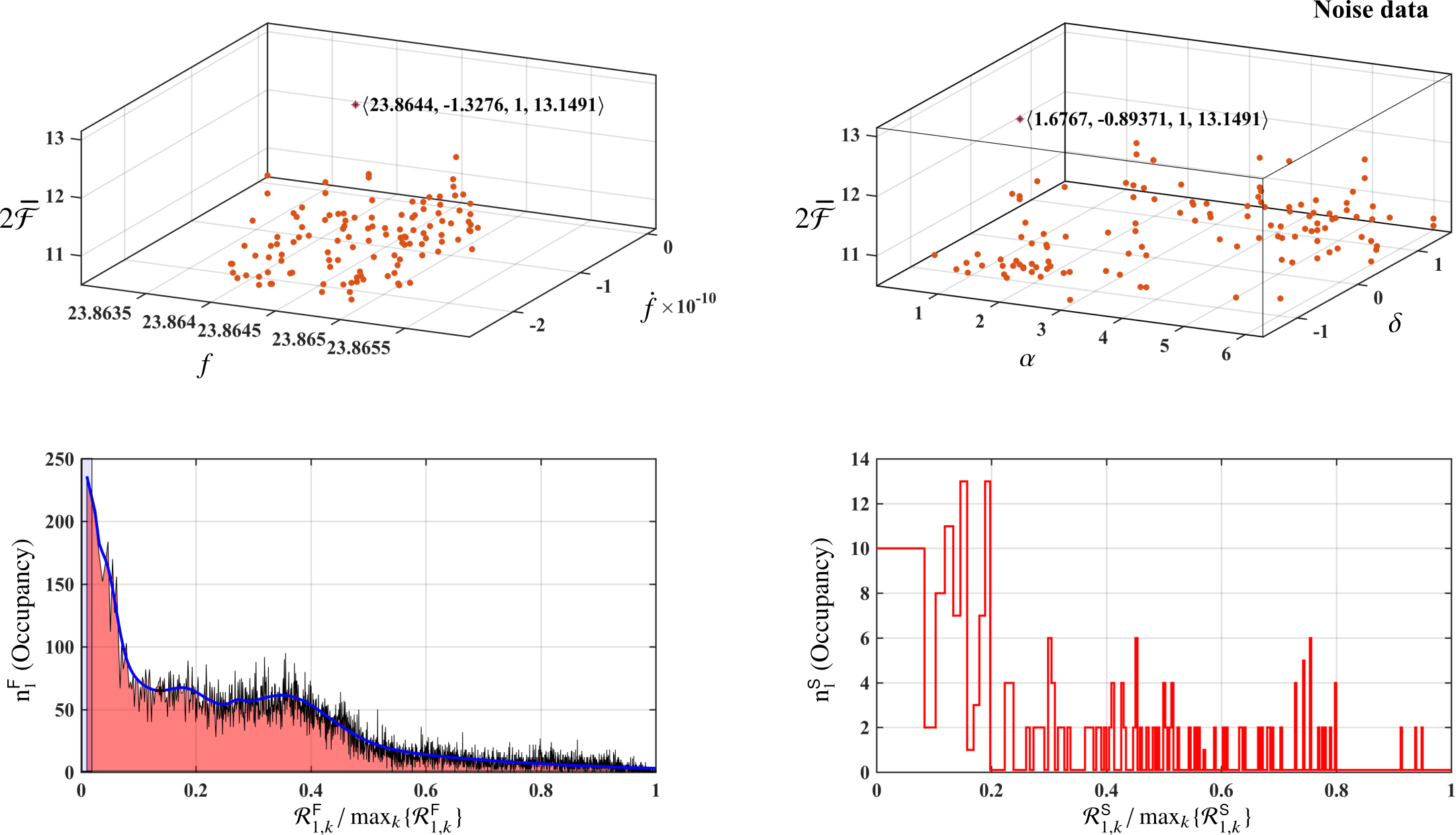}
\caption{(\tit{\textbf{Noise only}}) The orange points are the candidates in $\upchi_1^{\mathsf{sky}}$. The corresponding distributions for $\RFik$ and $\RSik$ are shown in the second row of plots. The shaded regions extend up to $\RFstar$ (left plot) and $\RSstar$ (right plot). Note that $\RFstar$ fails the hill parameters test and is reset to $\BFiOne$. The distribution in \Sspaces satisfies $\nSiOne \neq \max_r\{\nSir\}$ (i.e. no over-density is sky near the seed), so $\upphi_1$ is a single-occupant-cluster. The seed is marked with a `+'. The numbers in the brackets by the `+' denote: $\langle f, \dot{f}, i, \Gamma_i\rangle$ in \Fspaces and $\langle \alpha, \delta, i, \Gamma_i\rangle$ in \Sspaces, where $\Gamma\equiv \AvTwoF$.}
\label{fig:samplerun2}
\end{figure}

We now illustrate the different phases of the first iteration of the clustering procedure on two small snippets of data from the LIGO O1 run with and without a fake signal (figure \ref{fig:samplerun1}, figure \ref{fig:samplerun2}).
\end{multicols}

\end{document}